\documentclass[11pt]{article}
\usepackage{cite}
\usepackage{graphicx}
\usepackage{amssymb}
\advance\textheight by \topskip
\begin{document}
\tolerance=100000
\thispagestyle{empty}
\setcounter{page}{0}

\def\cO#1{{\cal{O}}\left(#1\right)}
\newcommand{\cp}{\hspace{1mm}\mbox{\raisebox{.3mm}{$\diagup$} \hspace{-7.3mm} C \hspace{-3.2mm} P}\hspace{1mm}}
\newcommand{\br}{\begin{eqnarray}}
\newcommand{\er}{\end{eqnarray}}
\newcommand{\bi}{\begin{itemize}}
\newcommand{\ei}{\end{itemize}}
\newcommand{\bn}{\begin{enumerate}}
\newcommand{\en}{\end{enumerate}}
\newcommand{\bc}{\begin{center}}
\newcommand{\ec}{\end{center}}
\newcommand{\ul}{\underline}
\newcommand{\ol}{\overline}
\newcommand{\ra}{\rightarrow}
\newcommand{\sm}{SM}
\newcommand{\as}{\alpha_s}
\newcommand{\aem}{\alpha_{em}}

\def\dis{\displaystyle}
\def\beq{\begin{equation}}
\def\eeq{\end{equation}}
\def\barr{\begin{array}}
\def\earr{\end{array}}

\def\U{{\cal U}}
\newcommand{\taulg}      {\ensuremath{\tau^{\pm} \to \ell^{\pm} \gamma}}

\def\Ecm{\ifmmode{E_{\mathrm{cm}}}\else{$E_{\mathrm{cm}}$}\fi}
\def\lsim{\buildrel{\scriptscriptstyle <}\over{\scriptscriptstyle\sim}}
\def\gsim{\buildrel{\scriptscriptstyle >}\over{\scriptscriptstyle\sim}}
\def \lum{{\cal L}}

\def\lapp{\mathrel{\rlap{\raise.5ex\hbox{$<$}}
                    {\lower.5ex\hbox{$\sim$}}}}
\def\gapp{\mathrel{\rlap{\raise.5ex\hbox{$>$}}
                    {\lower.5ex\hbox{$\sim$}}}}
\vspace*{1.0cm}
\begin{center}
{\Large \bf
Unparticles and Muon Decay}
 \\[0.30cm]
{\large  Debajyoti Choudhury$^{a}$, Dilip Kumar Ghosh$^{a}$ and 
Mamta$^{b}$}\\[0.35 cm]
$^a$Department of Physics and Astrophysics, University of Delhi, Delhi 110 007, India.\\[0.15cm]
$^b$Department of Physics, S.G.T.B. Khalsa College, 
University of Delhi, Delhi 110 007, India. 
\end{center}

\vspace{.2cm}

\begin{abstract}
Recently Georgi has discussed the possible
existence of `Unparticles' describable by operators having
non-integral scaling dimensions. With the interaction of these with
the Standard Model (SM) particles being constrained only by gauge and
Lorentz symmetries, it affords a new source for lepton flavour
violation. Current and future muon decay experiments are 
shown to be very sensitive to such scenarios.
\end{abstract}
\newpage

The notion of scale invariance in the description of a physical system
is a very powerful one and has found wide applications in many
different subdisciplines. A very well-known manifestation is afforded
by phase transitions wherein the existence of a critical temperature
is but a reflection of fluctuations at all length scales being equally
important.  In field theoretic models, scale invariance has
traditionally been a powerful tool in the analysis of the asymptotic
behaviour of correlation functions. And as is well known, conformal
invariance plays an even more fundamental role in string theories.

In the regime of particle physics though, the existence of many different 
particles (elementary or composite) with a very wide range of masses,  
manifestly breaks scale invariance. Indeed, an interacting scale invariant
theory in four space-time dimensions is, by definition, bereft of
particles thus running counter to our understanding of nature. 
Nonetheless, it is quite possible that there could exist a different 
sector of the theory that is so weakly coupled to the 
Standard Model (SM) particles that we have been unable to probe 
it experimentally. Clearly, this new physics is allowed to be 
described by a nontrivial scale invariant theory 
sector with an infrared fixed point. A concrete example is afforded 
by a vector-like non-abelian gauge theory with a large
number of massless fermions as studied by Banks and Zaks 
(BZ)~\cite{Banks:1981nn}. Supersymmetric nonlinear sigma 
models with similar features have also been considered 
in the literature~\cite{Braaten:1985is}.

In a recent paper, Georgi~\cite{Georgi:2007ek} investigated the
consequences of such a nontrivial scale invariant (BZ) sector
interacting with the SM fields through the exchange of (unspecified)
very heavy particles. Below the messenger scale, then, such
interactions between the BZ and the SM fields may be parametrized in
terms of nonrenormalizable interactions.  As scale-invariance in the
BZ sector emerges at an energy scale $\Lambda_\U$, this sector should
no longer be described in terms of conventional particles, but rather
in terms of massless ``unparticles''.  The renormalizable couplings of
the BZ fields cause a dimensional transmutation~\cite{coleman-weinberg}, 
and in the effective
theory operative below the scale $\Lambda_\U$, the BZ operators match
onto corresponding unparticle operators. The aforementioned
nonrenormalizable interactions, written in terms of unparticles,
would, in general, have non-integral scale dimensions, with very
unexpected phenomenological consequences~\cite{Georgi:2007ek,
Georgi:2007si, Cheung:2007ue, Luo:2007bq, Chen:2007vv, Ding:2007bm,
Liao:2007bx,Aliev:2007qw,Li-Wei,Murug,Cai-Dian,Stephanov,Fox}.

Since we have no direct information on either the unparticle or the
messenger sector, the only recourse for us in the exploration of the
interaction with the SM sector is to consider all possible operators
in an effective theory consistent with the symmetries of the SM.  In
particular, this includes flavour-changing
operators~\cite{Georgi:2007ek, Luo:2007bq}, and, more precisely, those
that violate lepton flavour conservation.

While lepton flavour violation (LFV) is absent in the minimal version
of the SM, it can be easily accommodated by extending the SM to
include neutrino masses. Indeed, the very observation of neutrino
oscillations~\cite{nu_osc} implies LFV. 
Including finite mass differences from neutrino mixing imply $\ell_i -
\ell_j$ mixing is generated at the one-loop level and is thus
suppressed by a factor of $(m^2_\nu/m^2_W)^2$.  However, various
extensions of the SM naturally incorporate large LFV effects.  The
simplest examples are afforded by the inclusion of heavy singlet Dirac
neutrinos~\cite{amon}, heavy right-handed Majorana neutrinos or
left-handed and right-handed neutral isosinglets~\cite{Cvetic:2002jy};
or even dimension-six effective fermionic operators~\cite{dim6}.  More
ambitious models consider see-saw mechanism with or without grand
unification~\cite{GUT}, supersymmetry~\cite{SUSY},
technicolor~\cite{techni}, models of compositeness~\cite{comp}.
Higgs~\cite{higgs} or a $Z'$-mediated~\cite{zpr} LFV has also been
considered in the literature. 

The great theoretical interest in LFV has been reflected in 
various experimental efforts as well. For example, each of the four 
LEP collaborations have investigated such scenarios at length~\cite{adlo}. 
This has also constituted an important component of the two 
collaborations at HERA~\cite{hera} and, perhaps more expectedly, at 
the two $B$-factories~\cite{Bfac}. And finally, several 
dedicated experiments have been designed to explore LFV. 
Of particular interest to us is the MEG experiment at the PSI~\cite{meg}, 
designed to detect forbidden decays of the muon down to the $10^{-14}$ 
level. 

Quite understandably, the study of muon decays have been a bedrock of 
the investigations into lepton flavour violation. Apart from the experimental 
ease, the very smallness of the muon decay width in the SM makes it
particularly amenable to look for small new physics effects. This is a 
feature that we wish to exploit. 

For unparticles, a discussion of LFV must necessarily be
attempted in the effective Lagrangian framework and hence part of it
bears resemblance to some of the above mentioned analyses, although
with very significant differences. Given our ignorance of the 
unparticle sector, all we can aver is that the unparticle operators in 
the effective Lagrangian must be SM gauge singlets and must have a 
mass dimension larger than one. They might have any Lorentz structure 
themselves as long as the overall operator is a Lorentz scalar. 

Given that the only decay mode allowed to the muon within the SM 
is that into an electron and missing energy-momentum 
($\mu^- \to e^- \, \bar \nu_e \, \nu_\mu$), the unparticle mode that could 
possibly fake it is $\mu^- \to e^- + \U$ and we shall start our analysis 
with this. Since the effective Lagrangian is perforce restricted 
to terms of the form 
\[
{\cal L} \supset {\cal O}^i_{SM} \; {\cal O_U}^i
\] 
where $i$ runs over Lorentz as well as flavour indices, 
the simplest term 
 that one can write for the process under consideration 
involves a scalar unparticle operator 
${\cal O_U}$ and can be expressed as 
\beq
{\cal L}_1 = 
\Lambda^{-d_u} \; {\bar e} \, \gamma_\eta \, (c_1+c_2 \, \gamma_5) 
\mu \; \partial^\eta {\cal O_U}  
    \label{coup_scalar}
\eeq
where $c_i$ are constants, $\Lambda \;(\equiv \Lambda_\U)$ is the
scale of new physics, and $d_u > 1 $ is the mass dimension of the
operator ${\cal O_U}$. Note that this Lagrangian had been considered
in Ref.\cite{Georgi:2007ek} in the context of the $t \to c + \U$ decay
wherein a choice $c_1 = - c_2 = 1$ was made for the analogous
coefficients. For reasons mentioned above, muon decay is expected to
be a far more sensitive probe of such couplings.

Using scale invariance to fix the two point correlators of the unparticle 
operators~\cite{Georgi:2007ek}, viz.
\beq
\barr{rcl}
\dis 
\left\langle0\right|{\cal O_U}(x)\,{\cal O_U}^\dagger(0)\left|0\right\rangle
& = & \dis 
\int\, \frac{d^4 p}{(2 \, \pi)^4} e^{-i \, P \cdot x}\,
\left|\left\langle0\right|{\cal O_U}(0)\left|P\,\right\rangle\right|^2\,
\rho\left(P^2\right)
\earr
\eeq
where $|P\rangle$ is the unparticle state of momentum $P^\mu$ created 
from the vacuum by the operator ${\cal O_U}$, and $\rho(P^2)$ is the 
density of states, we have~\cite{Georgi:2007ek}
\beq
\barr{rcl}
\dis 
\vert \langle 0 \vert O_\U (0)\vert P \rangle \vert^2
\rho(P^2)  & = & \dis
 A_{d_u}\, \theta(P^0) \, \theta (P^2) \,(P^2)^{d_u -2 } \ ,
\\[2ex]
{\rm with} \qquad    A_{d_u} & \equiv & \dis 
\frac{16 \, \pi^{5/2} } { (2 \, \pi)^{2 \, d_u} }  \;
       \frac{ \Gamma({d_u}+{1 \over
       2})}{\Gamma({d_u}-1) \; \Gamma(2\,{d_u})}   
\earr
\eeq
normalised to give the phase space for $d_u$ massless particles. 
The decay profile can then be computed in a straightforward manner to 
yield~\cite{Georgi:2007ek}
\beq
\barr{rcl}
\dis 
\frac{d\Gamma_S}{d E_e} (\mu \to e + {\cal U}) & = & 
\dis \frac{A_{d_u}}{4 \, \pi^2} \; (c_1^2 + c_2^2) \; 
m_\mu^2 \; E_e^2 \; \left(m_\mu^2 - 2 \, m_\mu \, E_e \right )^{d_u-2} 
\Lambda^{-2 \, d_u} \; \Theta(m_\mu - 2 \, E_e)
\\[2ex]
\Gamma_S (\mu \to e + {\cal U}) 
     & = & \dis \frac{A_{d_u}}{16 \, \pi^2} \; \frac{c_1^2 + c_2^2}{d_u^3 -d_u }
     \; 
m_\mu  \; \left(\frac{m_\mu}{\Lambda}\right)^{2 \, d_u} 
\earr
     \label{decay_scalar}
\eeq
where the mass of the electron has been neglected and the second equality
follows only for $d_u > 1$. 
\begin{figure}[!h]
\begin{center}
\vspace*{-0.2cm}
\includegraphics[width= 10 cm, height= 8 cm]{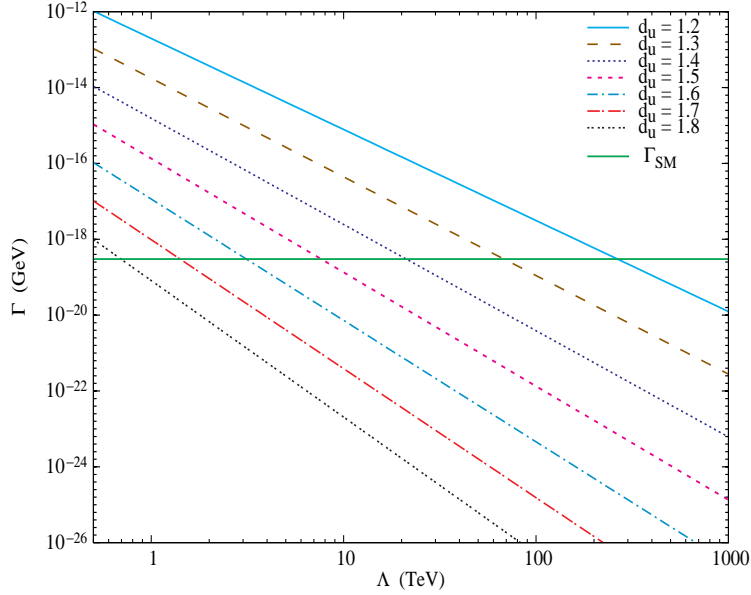}
\vspace*{-0.4cm}
\caption{\em The muon decay width into ($e^- + \U$) as a function of 
the unparticle physics scale $\Lambda$ for various values of the 
mass dimension $d_u$ of the scalar operator ${\cal O_U}$. 
We have adopted the convention $c_1^2 + c_2^2 = 1$. Also shown is the 
SM width for the muon.
}
\label{fig:scalar_width}
\end{center}
\end{figure}

\begin{figure}[!h]
\begin{center}
\vspace*{-0.2cm}
\includegraphics[width= 10 cm, height= 8 cm]{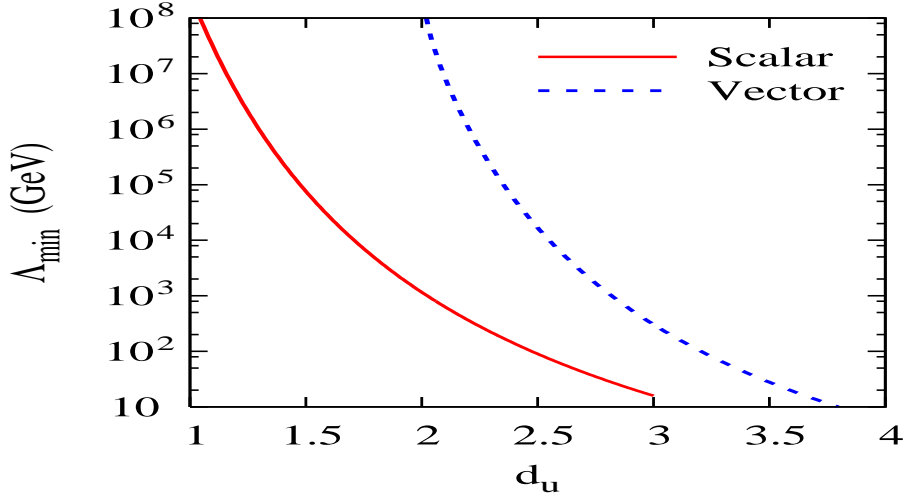}
\vspace*{-0.4cm}
\caption{\em The lower limit on the scale of the effective theory 
for scalar (${\cal O_U}$) and vector(${\cal O_U^\eta}$) operators
as a function of their mass dimension and assuming 
that $Br(\mu^- \to e^- + \U) \leq 10^{-3}$. In either case, we 
have assumed that $c_1^2 + c_2^2 = 1$ and $c_3^2 + c_4^2 = 1$ respectively.}
\label{fig:limits}
\end{center}
\end{figure}

In Fig.\ref{fig:scalar_width}, we display the total width as a function of
$\Lambda$ for different choices of $d_u$. Since the dependence on the 
coefficients $c_i$ is trivial, we have made the simplifying assumption of 
\[
    c_1^2 + c_2^2 = 1 \ 
\]
(a convention often adopted in effective field theories). To draw
conclusions about the sensitivity of this measurement to unparticle 
physics, it needs be remembered that muon decay is one of the best 
measured observables and, in fact, essentially constitutes the 
measurement of the Fermi coupling constant~\cite{pdg,g_mu}. Furthermore, 
$G_F$ is an input for various other precision measurements, a notable one 
being the coupling of the $W$-boson to the first generation 
quarks~\cite{pdg,vud}, 
viz,
\[
   V_{u d} = 0.97377 \pm 0.00027 \ .
\]
Since the latter is determined from superallowed nuclear beta decays, 
the couplings $c_{1,2}$ have no r\^ole to play here. Thus, 
barring magical conspiracies between different terms in the effective 
Lagrangian, we may safely demand
\beq
     \Gamma(\mu^- \to e^- + \U) \leq 
       10^{-3} \; \Gamma(\mu^- \to e^- + \bar\nu_e + \nu_\mu) \ ,
\eeq
and the consequent bounds are presented in Fig.\ref{fig:limits}. 
These, expectedly,  are quite strong, especially for small $d_u$.  
And, while these have been derived for $c_1^2 + c_2^2 = 1$, 
the dependence is quite mild, with 
the bounds obtained on $\Lambda$ scaling  
as $(c_1^2 + c_2^2)^{1 / 2 \, d_u}$. 

\begin{figure}[!h]
\begin{center}
\vspace*{-0.2cm}
\includegraphics[width= 10 cm, height= 8 cm]{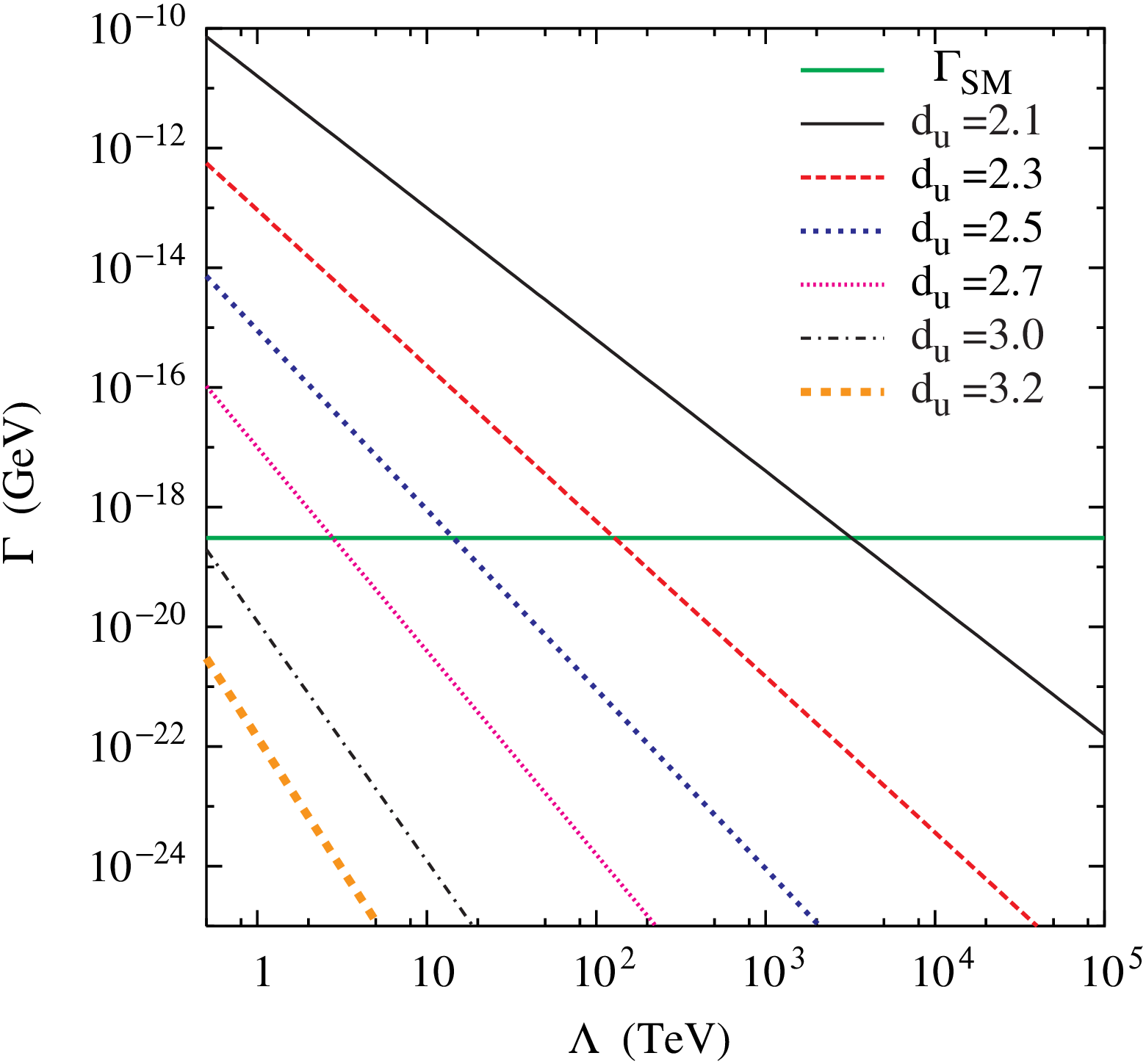}
\vspace*{-0.4cm}
\caption{\em The muon decay width into ($e^- + \U$) as a function of 
the unparticle physics scale $\Lambda$ for various values of the 
mass dimension $d_u$ of the vector operator ${\cal O_U^\eta}$. 
We have adopted the convention $c_3^2 + c_4^2 = 1$. Also shown is the 
SM width for the muon.
}
\label{fig:vector_width}
\end{center}
\end{figure}

We now consider a different possible coupling of the 
unparticles to the muon-electron current, namely a vector one:
\beq
{\cal L}_2  = 
\Lambda^{1-d_u} \; {\bar e} \, \gamma_\eta \, (c_3 +c_4 \, \gamma_5) 
\mu \; {\cal O_U^\eta}
\eeq
where ${\cal O_U^\eta}$ is a transverse and Hermitian operator. 
The transversality condition, alongwith scale invariance, now 
stipulates that 
\beq
\left\langle0\right|{\cal O_U^\eta}(0)\left|P\,\right\rangle\,
\left\langle P\right|{\cal O_U^\omega}(0)\left|0\,\right\rangle\,
\rho\left(P^2\right)
=A_{d_u}\,\theta\left(P^0\right)
\,\theta\left(P^2\right)\,
\left(-g^{\eta \omega}+P^\eta P^\omega/P^2\right)
\,\left(P^2\right)^{d_u -2} \ .
    \label{polsum}
\eeq
This leads to 
\beq
\barr{rcl}
\dis \frac{d\Gamma_V}{d E_e} (\mu \to e + \U) & = & \dis
\frac{A_{d_u}}{4 \, \pi^2} \; (c_3^2 + c_4^2)  \; 
m_\mu \; E_e^2 \, \left(m_\mu^2 - 2 \, m_\mu \, E_e \right )^{d_u-3} 
\Lambda^{2-2 \, d_u} \, (3 m_\mu - 4 E_e) \;
\Theta(m_\mu - 2 \, E_e)
\\[2ex]
\Gamma_V (\mu \to e + \U) & = & \dis 
\frac{3 \, A_{d_u}}{16 \, \pi^2} \; \frac{c_3^2 + c_4^2}{d_u^3 - d_u^2 - 2 \, d_u}
  \; m_\mu \; \left( \frac{m_\mu}{\Lambda} \right)^{2 \, d_u - 2}
\earr
     \label{decay_vector}
\eeq 
once again neglecting the electron mass. The second equality 
holds only for $d_u > 2$. The resultant total width 
is displayed in Fig.\ref{fig:vector_width}. 

Certain differences with the
scalar case (Eq.\ref{decay_scalar}) are easy to appreciate. The
coupling of the fermion current to the scalar operator ${\cal
O_U}$---Eq.\ref{coup_scalar}---is a helicity suppressed one, leading
to the amplitude being proportional to $m_\mu$. With the coupling to
the vector operator being free of this suppression, one would naively
expect an enhancement, relative to the scalar case, by roughly a
factor of $(\Lambda / m_\mu)^2$.
In other words, the constraints on $\Lambda$, for identical values of
$d_u$, are expected to be much stronger for the vector case than that
for the scalar one. That this is indeed true can be easily divined 
from a comparison of Figs.\ref{fig:scalar_width}\&\ref{fig:vector_width}.

On the other hand, note that 
the differential width is now proportional to 
$\left(m_\mu^2 - 2 \, m_\mu \, E_e \right )^{d_u-3}$, or, in other 
words, has an extra factor of $1 / P^2$. 
This, of course, can be traced to 
Eq.\ref{polsum}. While this term would not contribute when ${\cal O_U^\eta}$ 
couples to a conserved current, in the present context it leads to 
an enhanced density of states in the small $P^2$ regime. Consequently, 
the total width is divergent unless $d_u > 2$. This constitutes a key 
result of our study and we shall return to it later. 

The limits on the effective scale for vector-like couplings
are displayed in Fig.\ref{fig:limits}. Once again, we have assumed 
that $c_3^2 + c_4^2 = 1$. While it may seem that the limits are 
much stronger for the  vector case, note that, given the 
structure of ${\cal L}_1$ and ${\cal L}_2$, it is only fair to 
compare $\Gamma_V(d_u, \Lambda)$ with  $\Gamma_s(d_u - 1, \Lambda)$. 
Shifting the curve for the vector coupling in Fig.\ref{fig:limits} 
to the left by one unit shows that the new curve would, for the most part, 
fall below that for the scalar. This can be easily understood by 
considering the ratio
\[
\dis
\frac{\Gamma_S(d_u - 1, \Lambda)}{\Gamma_V(d_u , \Lambda)}
 =  \frac{16 \, \pi^2}{3} \; \frac{c_1^2 + c_2^2}{c_3^2 + c_4^2} \;
      (d_u - 2) \, (d_u + 1)  \qquad \quad (d_u > 2)
\]
which is larger than unity unless $d_u$ is very close to 2. 

\begin{figure}[!h]
\begin{center}
\vspace*{-0.2cm}
\includegraphics[width= 10 cm, height= 8 cm]{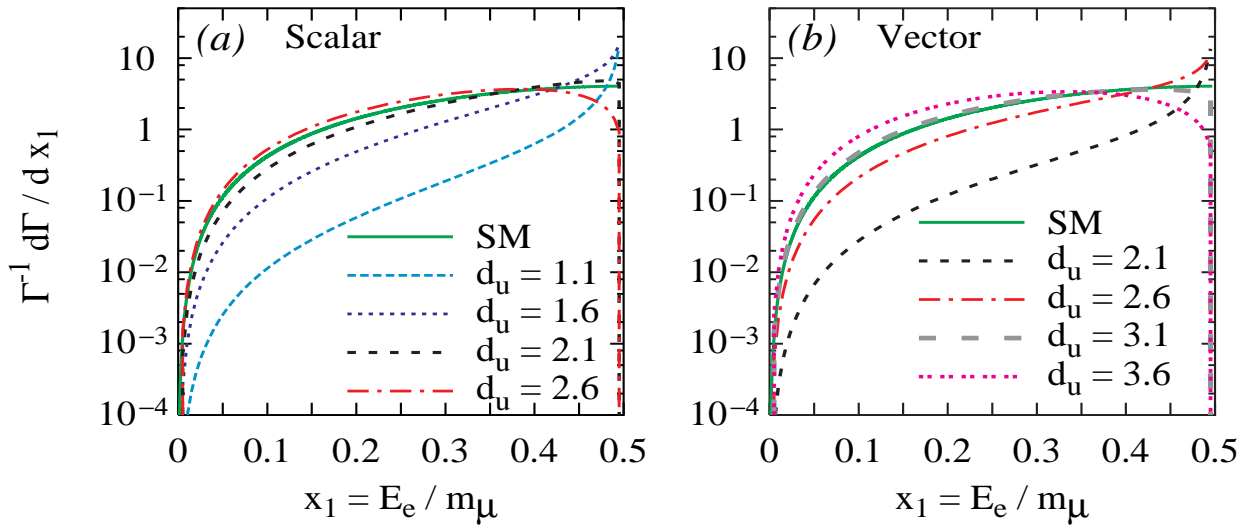}
\vspace*{-0.4cm}
\caption{\em The energy distribution for the electron in 
$\mu^- \to e^- + \U$ decay for various values of $d_u$. The 
left (right) panels refer to scalar (vector) unparticle operators 
respectively. Also shown, in each case, is the 
energy distribution for the SM process $\mu^- \to e^- + \bar \nu_e + \nu_\mu$.
}
\label{fig:dsigma_scal_vec}
\end{center}
\end{figure}

It is amusing to consider the hypothetical case of an observed discrepancy 
in the decay $\mu^- \to e^- + {\rm nothing}$ in the forthcoming 
experiments. For example, can  MEG~\cite{meg} distinguish between the 
possible unparticle operators if such deviation were to be seen? A possible
means is provided by the shape of the energy distribution. In 
Fig.\ref{fig:dsigma_scal_vec}, we display the same for both cases considered
above. 
For small values of $d_u$, the distributions are naturally peaked at 
$E_e = m_\mu / 2$ as is expected for a decay into two massless particles. 
While it might seem that, for 
the vector case, the peaking persists to much larger values of $d_u$, it 
is but a reflection of the differing powers of $P^2$ in the two 
cases ($d_u - 3$ for vector vs. $d_u - 2$ for scalar). 

It should be noted here that, while the limit $d_u \to 1^+$ 
for the scalar case 
corresponded to the two-body decay, in the case of the vector, it is instead 
the limit $d_u \to 2^+$ that corresponds to the same (namely, a muon decaying 
to an electron and a vector particle). Similarly, $d_u \to 2$ in the scalar 
case corresponds to a three-body decay (and hence the close identification 
with the SM curve in Fig.\ref{fig:dsigma_scal_vec}$a$). For the vector case, 
this feature is exhibited in the $d_u \to 3$ limit instead. Both these 
correspondences in the vector case owe themselves to the form of 
Eq.\ref{polsum} and are reflective of the fact that, in this case, 
it is $d_u \to 2^+$ that goes over to the one-particle description and 
hence the theory makes sense only for $d_u > 2$. 

Until now, we have been considering only LFV couplings of the unparticle 
sector with SM matter. Of course, this sector could couple to lepton flavour 
conserving currents as well. As far as muon decays are concerned, the only 
such coupling that is of relevance is the one with the electrons. Restricting 
ourselves to the vector operator, we may now write an additional term of the 
form 
\beq
{\cal L}_3 = 
\Lambda^{1-d_u} \; {\bar e} \, \gamma_\eta (c_5+ c_6 \, \gamma_5)  \, 
e \; {\cal O_U^\eta}
\eeq
Such terms would immediately manifest themselves in 
observables pertaining to the electron, in particular low-energy ones. 
In Ref.\cite{Liao:2007bx}, effects on
both the anomalous magnetic moment of the electron and the decays 
of ortho-positronium was examined. The bounds were found to be 
quite stringent, in particular those emanating from the latter set of 
observables. It should be noted that unparticle contributions to 
such observables are not 
proportional to the combination $(c_5^2 + c_6^2)$. 
For example, a value $d_u = 1.5$ would imply 
$\Lambda / c_5 \geq 4.3 \times 10^5$ TeV, or 
$\Lambda / c_6 \geq 510$ TeV, as long as only one of the two 
coefficients were to be non-zero\footnote{Note that the use of $d_u < 2$ 
may also be a cause for concern in this context.}. 
Of course, large cancellations 
between several such contributions are possible, but would represent 
a fine-tuned situation. 

Simultaneous presence of both sets of operators ${\cal L}_2$ and 
${\cal L}_3$  would immediately engender unparticle-mediated 
$\mu \to 3 \, e$ decays. The calculation is straightforward and mirrors 
that in the presence of a LFV $Z'$. Since the vector propagator in this 
case is given by~\cite{Georgi:2007si,Cheung:2007ue}
\[
\begin{array}{c}
\displaystyle
\int\,e^{iPx}\,
\left\langle0\right|T({\cal O_U^\mu}(x)\,{\cal O_U^\nu}(0))
       \left|0\right\rangle\,d^4x
=
\frac{i}{2} \, A_{d_u}\,
\frac{-g^{\mu\nu}+P^\mu P^\nu/P^2}{\sin(d_u \, \pi)}\,
\left(-P^2-i\epsilon\right)^{d_{u}-2} \ ,
\end{array}
\]
the spin-summed and averaged matrix-element-squared for the decay
$\mu (p_1) \to e^-(p_2) + e^-(p_3) + e^+(p_4) $ can be computed 
to be
\beq
\barr{rcl}
\dis 
\left [\frac{4\Lambda^{4-4 d_u} A^2_{d_u}}{\sin^2(d_u\pi)} \right ]^{-1} 
\overline{\mid {\cal M}\mid^2} &=&  \dis 
\left|{\cal P}_1\right|^2 \;
\left[K_1 \, (p_{23} \, p_{14} + p_{24} \, p_{13}) + K_2 \, 
             (p_{23} \, p_{14} - p_{24} \, p_{13})\right] 
\\[2ex]
& + & \left|{\cal P}_1\right|^2 \;
\left[K_1 \, (p_{23} \, p_{14} + p_{34} \, p_{12}) 
    + K_2 \, (p_{23} \, p_{14} - p_{34} \, p_{12})\right] 
\\[2ex]
&- & \dis 2 \, Re\left( {\cal P}_1 \; {\cal P}_2^* \right) \;
\left[K_1 + K_2\right] \, (p_{14} \, p_{23})
\\[2ex]
{\cal P}_1 & \equiv &  \left[- (p_1 - p_2)^2 - i \, \epsilon\right]^{d_u - 2} 
\\[2ex]
{\cal P}_2 & \equiv &  \left[- (p_1 - p_3)^2 - i \, \epsilon\right]^{d_u - 2} 
\\[2ex]
 K_1 & \equiv & \dis \left( c_3^2 + c_4^2 \right) \;
                     \left( c_5^2 + c_6^2 \right) 
\\[1ex]
K_2 & \equiv & 4 \, c_3 \, c_4 \, c_5 \, c_6 
\earr
     \label{mu_3e}
\eeq where $p_{ij} \equiv p_i. p_j $, and we have suppressed terms of
${\cal O}(m_e)$ in the first equation for reasons of brevity\footnote
{ This process is also discussed in Ref.\cite{Aliev:2007qw}. However,
they concentrate on $d_u <2$, a regime that is unphysical.}.
Integrating Eq.(\ref{mu_3e}) over the phase space would give us the
total partial width in this channel. It should be noted though that
considering strictly massless electrons would lead to a divergent
value of the matrix-element whenever the positron were to be collinear
with either of the electrons. Although it is numerically sufficient to
consider the phase space to be that for three massive particles, while
continuing to neglect $m_e$ in the first of Eqs.(\ref{mu_3e}), in our
calculations, we retain the full dependence on $m_e$.

\begin{figure}[!h]
\begin{center}
\vspace*{-0.2cm}
\includegraphics[width= 10 cm, height= 8 cm]{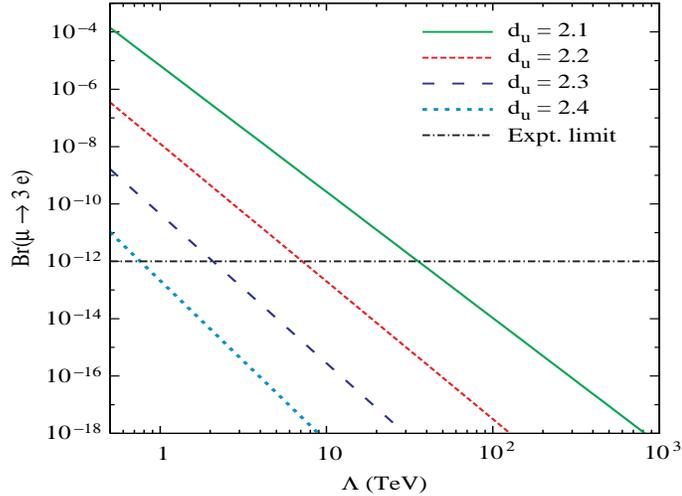}
\vspace*{-0.4cm}
\caption{\em $Br(\mu \to 3 \, e)$ as a function of 
the unparticle physics scale $\Lambda$ for various values of the 
mass dimension $d_u$ of the vector operator ${\cal O_U^\eta}$. Both 
${\cal L}_2$ and ${\cal L}_3$ terms appear in the effective Lagrangian
with $c_3 = c_4 = c_5 = c_6 = 1/\sqrt{2}$. 
Also shown is the experimental upper limit for this channel.}
\label{fig:mu3e}
\end{center}
\end{figure}

In Fig.~\ref{fig:mu3e}, we present the branching ratio 
$Br(\mu \to 3 \, e)$ as a function of the scale $\Lambda$ 
for different values of the scaling dimension $d_u$. To 
facilitate easy comparison with the limits obtained earlier, 
we maintain $c_3 = c_4 = c_5 = c_6 = 1/\sqrt{2}$. We concentrate 
on $d_u > 2$ here for the aforementioned reasons. Although the 
experimental limit is~\cite{pdg}
\[
Br (\mu \to 3 \, e) < 10^{-12} \ ,
\]
note that the constraints on $\Lambda$ from this process are 
typically much weaker than those already obtained from 
$\mu \to e + \U$. The reason is not far to seek. Compared to 
the 2-body decay, the rate for $\mu \to 3 \, e$ 
process involves an extra factor of $(m_\mu / \Lambda)^{2 \, d_u - 2}$, 
apart from phase space factors and, consequently, the rate is suppressed 
even for the smallest of $d_u$ allowed. It might seem then that 
it is pointless to consider $\mu \to 3 \, e$, given the already existent 
constraints. But before we conclude so, it is of importance to ask whether 
$\mu^- \to e^- + \U$ could fake $\mu^- \to e^- \, \bar v_e \, \nu_\mu$ 
even in the presence of sizable $c_{5, 6}$. Although it has been 
argued~\cite{Cheung:2007ue}
that the imaginary part of the unparticle propagator does not correspond 
to a finite decay width, and that the unparticle, once produced, never decays.
Remember though that the entire formalism corresponds to an effective theory
and the details lie in the ultraviolet completion.  
In a very recent deconstruction of this theory, Stephanov~\cite{Stephanov}
points out that the unparticle can be viewed as the limiting case of 
an infinite tower of particles of
different masses with a regular mass spacing. If the spacing is small, 
but finite, then the unparticles are allowed to decay. In view of such 
subtleties and the lack of knowledge on our part as to the exact 
nature of unparticles (were they to be discovered), it seems 
contingent upon us to explore each constraint on its own.

\begin{figure}[!h]
\begin{center}
 \vspace*{1.5cm}
\includegraphics[width= 10 cm, height= 10 cm]{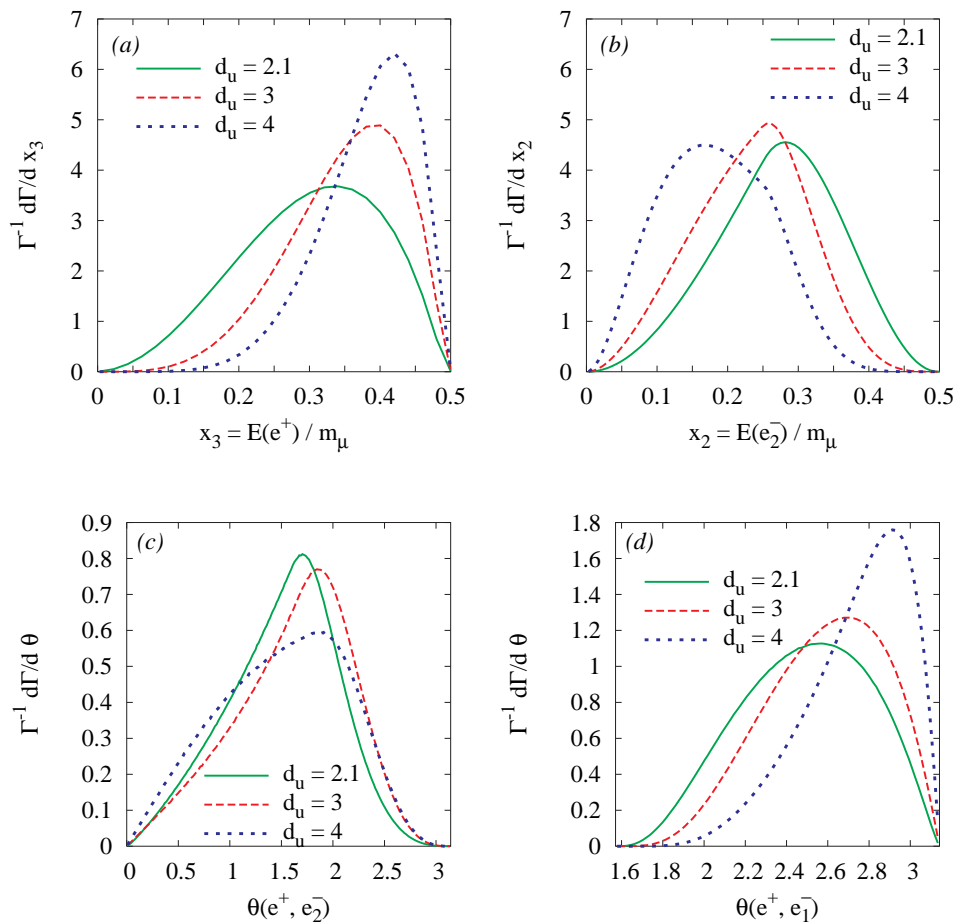}
\vspace*{1.0cm}
\caption{\em Various normalized phase space distributions (in the 
muon rest frame) 
for the decay $\mu^- \to e^+ \, e^- \, e^-$ mediated by vector unparticles. 
{\em (a)} the energy of the positron; 
{\em (b)} the energy of the softer electron;
{\em (c)} the angle the softer electron subtends with the 
          positron;
{\em (d)} the angle the harder electron subtends with the 
          positron.
}
\label{fig:mu3e_diff}
\end{center}
\end{figure}

It is both amusing and instructive to consider the phase 
space distributions for the $\mu^- \to e^+ \, e^- \, e^-$ decay. 
Concentrating, for simplicity, on unpolarized muons, 
we present, in Fig.\ref{fig:mu3e_diff}, some of these
distributions in the muon rest frame. The dependence on $d_u$ 
is  quite striking. For $d_u = 3$, each of these (and any other)
matches the corresponding distributions for say a $Z'$-mediated 
$\mu \to 3 \, e$ decay. This, of course, is expected since 
$d_u = 3$ corresponds to a single vector exchange. For $d_u > 3 \; (< 3)$, 
the positron spectrum becomes harder (softer), while the reverse is true of 
the softer of the two electrons. Similarly, $d_u > 3 \; (< 3)$ 
pushes the softer of the two electrons farther (closer) to the positrons. 

The simultaneous presence of both ${\cal L}_2$ and ${\cal L}_1$ would also 
lead to processes like $e^+ + e^- \to \mu^+ + e^-$ at high energy 
colliders and presumably used to look for 
unparticle effects at linear colliders.
The amplitude for this can be obtained trivially 
by the use of crossing symmetry. A simple estimate shows though 
that, given the strong constraints already obtained, a
first generation linear collider would not add to the sensitivity.

To summarize, we have studied a particularly intriguing
aspect of low energy 
phenomena associated with Unparticle physics, namely nonconservation 
of lepton flavour. As unparticles are associated with a hidden scale invariant 
sector that communicates with the SM fields through a heavy messenger sector, 
at low energies such interactions are 
parametrized by generic operators 
in an effective field theory consistent with the symmetries of the 
SM. Of particular relevance here is the 
non-integral value of the 
scale dimensions of these operators, which could lead to very interesting
phenomenology. 

With lepton flavour violation being absent in the SM, 
it proffers an ideal theatre to look for signatures of physics 
beyond the SM. It is well known that, in the SM,
the only decay mode allowed to the muon is that into an
electron and missing energy-momentum $(\mu^- \to e^- {\bar{\nu_e}} \nu_\mu)$;
this channel could possibly be mimicked by $\mu^- \to e^- + \U $, where,
the missing energy-momentum is carried by the Unparticle $\U$.  While
the scale dimension $d_u > 1$ for the scalar operator, 
we demonstrate that consistency of the vector operator requires 
the corresponding scaling dimension to be greater than 2.
The present experimental accuracies on $G_\mu$ and the nuclear beta decay 
measurements lead to very strong bounds on the 
unparticle scale $\Lambda$. In case a deviation is 
observed in future muon decay experiments, we demonstrate 
how the shape of the electron energy 
distribution could disentangle unparticle 
effects from other possible electroweak physics. 

In addition to the real emission of unparticles, we reexamine $\mu \to
3 e$ decay mediated by a vector unparticle operator and find some 
disagreements with Ref.\cite{Aliev:2007qw}. 
And although the 
constraints obtainable from present upper bounds on this decay mode 
are weaker than those derivable from $\mu \to e + \U$, this mode does 
offer an oppurtunity to probe some interesting issues, both theoretical 
and experimental.

\section*{Acknowledgments}
The authors thank Kingman Cheung for 
useful discussions and the IUCAA Reference Centre, Delhi 
for infrastructural support. 
DC acknowledges support from the 
Dept. of Science and Technology, India under project number
SR/S2/RFHEP-05/2006. DKG and Mamta would respectively 
like to thank the Helsinki Institute 
of Physics, University of Helsinki  and the Center for HIgh
Energy Physics, Indian Institute of Science 
for hospitality while part of this 
work was completed. 



\begin{thebibliography}{99}

\bibitem{Banks:1981nn}
  T.~Banks and A.~Zaks,
  Nucl.\ Phys.\  B {\bf 196}, 189 (1982).


\bibitem{Braaten:1985is}
  E.~Braaten, T.~L.~Curtright and C.~K.~Zachos,
  Nucl.\ Phys.\  B {\bf 260}, 630 (1985).


\bibitem{Georgi:2007ek}
  H.~Georgi,
  arXiv:hep-ph/0703260.


\bibitem{Georgi:2007si}
  H.~Georgi,
  arXiv:0704.2457 [hep-ph].

\bibitem{coleman-weinberg}
S.~Coleman and E.~Weinberg, Phys. Rev. {\bf D}7, 1888 (1973).

\bibitem{Cheung:2007ue}
  K.~Cheung, W.~Y.~Keung and T.~C.~Yuan,
  arXiv:0704.2588 [hep-ph].


\bibitem{Luo:2007bq}
  M.~Luo and G.~Zhu,
  arXiv:0704.3532 [hep-ph].


\bibitem{Chen:2007vv}
  C.~H.~Chen and C.~Q.~Geng,
  arXiv:0705.0689 [hep-ph].

\bibitem{Ding:2007bm}
  G.~J.~Ding and M.~L.~Yan,
  arXiv:0705.0794 [hep-ph].

\bibitem{Liao:2007bx}
  Y.~Liao,
  arXiv:0705.0837 [hep-ph].


\bibitem{Aliev:2007qw}
  T.~M.~Aliev, A.~S.~Cornell and N.~Gaur,
  arXiv:0705.1326 [hep-ph].

\bibitem{Li-Wei}
X.~Q.~Li and Z.~T.~Wei, arXiv:0705.1821~[hep-ph].
\bibitem{Murug}
M.~Duraisamy, arXiv:0705.2622~[hep-ph]. 
\bibitem{Cai-Dian}
Cai-Dian Lu, Wei Wang and Yu-Ming Wang, arXiv:0705.2909 [hep-ph].
\bibitem{Stephanov}
M.~A.~Stephanov, arXiv:0705.3049 [hep-ph].
\bibitem{Fox}
P.~J.~Fox, A.~Rajaraman and Y.~Shirman, arXiv:0705.3092 [hep-ph].

\bibitem{nu_osc}
B.~T.~Cleveland {\it et al.}, 
Astrophys.\ J.\  {496} (1998) 505;
Y.~Fukuda {\it et al.}, Super-Kamiokande Collaboration,
Phys.\ Rev.\ Lett.\  {81} (1998) 1562;
Q.~R.~Ahmad {\it et al.},  SNO Collaboration,
Phys.\ Rev.\ Lett.\  {89} (2002) 011301;
M.~H.~Ahn {\it et al.},  K2K Collaboration,
Phys.\ Rev.\ Lett.\  {90} (2003) 041801.

\bibitem{amon}
A.~Ilakovac,
Phys.\ Rev.\ D {\bf 62}, 036010 (2000).

\bibitem{Cvetic:2002jy}
G.~Cvetic, C.~Dib, C.~S.~Kim, and J.~D.~Kim,
Phys.\ Rev.\ D {66} (2002) 034008;
[Erratum-ibid.\ D {68} (2003) 059901].

\bibitem{dim6}
D.~Black {\it et al.},
Phys.\ Rev.  D {\bf  66}, 053002 (2002).


\bibitem{GUT} 
J.~R.~Ellis, M.~E.~Gomez, G.~K.~Leontaris, S.~Lola, and D.~V.~Nanopoulos,
Eur.\ Phys.\ J.\ C {14} (2000) 319;\\
J.~R.~Ellis, J.~Hisano, M.~Raidal, and Y.~Shimizu,
Phys.\ Rev.\ D {66} (2002) 115013; \\
 A.~Masiero, S.~K.~Vempati, and O.~Vives,
Nucl.\ Phys.\ B {649} (2003) 189; \\
 T.~Fukuyama, T.~Kikuchi, and N.~Okada,
 Phys.\ Rev.\ D {68} (2003) 033012.

\bibitem{SUSY} 
A.~Dedes, J.~R.~Ellis, and M.~Raidal,
Phys.\ Lett.\ B {549} (2002) 159;\\
A.~Brignole and A.~Rossi,
Phys.\ Lett.\ B {566} (2003) 217;\\
J.P.~Saha and A.~Kundu,
Phys.\ Rev.\ D {\bf 66}, 054021 (2002);\\
R.~Barbier {\it et al.},
Phys.\ Rep.\ {\bf 420}, 1 (2005).

\bibitem{techni} S.~Dimopoulos and L.~Susskind, Nucl. Phys.  B~{\bf 155}
  (1979) 237;\\
  S.~Dimopoulos, Nucl. Phys.  B~{\bf 168} (1980) 69;\\
  E.~Farhi and L.~Susskind, Phys. Rev. D~{\bf 20} (1979) 3404;\\
  E.~Farhi and L.~Susskind, Phys. Rep. {\bf 74} (1981) 277;\\
C.~x.~Yue, Y.~m.~Zhang, and L.~j.~Liu,
Phys.\ Lett.\ B {547} (2002) 252.

\bibitem{comp} B.~Schrempp and F.~Schrempp, Phys. Lett. B~{\bf 153}
  (1985) 101;\\
  J. Wudka, Phys. Lett. B~{\bf 167} (1986) 337.


\bibitem{higgs}
M.~Sher,
Phys.\ Rev.\ D {\bf 66}, 057301 (2002);\\
A.~Brignole and A.~Rossi,
Nucl.\ Phys.\ B {\bf 701}, 3 (2004);\\
C.-H.~Chen and C.-Q.~Geng,
Phys.\ Rev.\ D {\bf 74}, 035010 (2006).



\bibitem{zpr}
W.-J.~Li, Y.-D.~Yang and X.-D.~Zhang,
  Phys.\ Rev.\ D {\bf 73}, 073005 (2006).


\bibitem{adlo}
  A.~Heister {\it et al.}  [ALEPH Collaboration],
  Eur.\ Phys.\ J.\  C {\bf 31}, 1 (2003)
  [arXiv:hep-ex/0210014]; \\
%
  A.~Heister {\it et al.}  [ALEPH Collaboration],
  Eur.\ Phys.\ J.\  C {\bf 25}, 1 (2002)
  [arXiv:hep-ex/0201013]; \\
%
  R.~Barate {\it et al.}  [ALEPH Collaboration],
  Eur.\ Phys.\ J.\  C {\bf 19}, 415 (2001)
  [arXiv:hep-ex/0011008]; \\
%
  R.~Barate {\it et al.}  [ALEPH Collaboration],
  Eur.\ Phys.\ J.\  C {\bf 13}, 29 (2000);\\
%
  R.~Barate {\it et al.}  [ALEPH Collaboration],
  Eur.\ Phys.\ J.\  C {\bf 4}, 433 (1998)
  [arXiv:hep-ex/9712013]; \\
%
  C.~Martinez-Rivero  [DELPHI Collaboration],
{\it Prepared for 28th International Conference on High-energy Physics (ICHEP 96), Warsaw, Poland, 25-31 Jul 1996}; \\
%
  P.~Abreu {\it et al.}  [DELPHI Collaboration],
  Phys.\ Lett.\  B {\bf 487}, 36 (2000)
  [arXiv:hep-ex/0103006]; \\
%
  P.~Abreu {\it et al.}  [DELPHI Collaboration],
  Phys.\ Lett.\  B {\bf 502}, 24 (2001)
  [arXiv:hep-ex/0102045]; \\
%
  P.~Abreu {\it et al.}  [DELPHI Collaboration],
  Eur.\ Phys.\ J.\  C {\bf 13}, 591 (2000); \\
%
  P.~Abreu {\it et al.}  [DELPHI Collaboration],
  Z.\ Phys.\  C {\bf 73}, 243 (1997);\\
%
  P.~Achard {\it et al.}  [L3 Collaboration],
  Phys.\ Lett.\  B {\bf 524}, 65 (2002)
  [arXiv:hep-ex/0110057]; \\
%
  M.~Acciarri {\it et al.}  [L3 Collaboration],
  Eur.\ Phys.\ J.\  C {\bf 19}, 397 (2001)
  [arXiv:hep-ex/0011087];\\
%
  M.~Acciarri {\it et al.}  [L3 Collaboration],
  Phys.\ Lett.\  B {\bf 459}, 354 (1999); \\
%
  B.~Adeva {\it et al.}  [L3 Collaboration],
  Phys.\ Lett.\  B {\bf 271}, 453 (1991);\\
%
  G.~Abbiendi {\it et al.}  [OPAL Collaboration],
  Phys.\ Lett.\  B {\bf 519}, 23 (2001)
  [arXiv:hep-ex/0109011];\\
%
  G.~Abbiendi {\it et al.}  [OPAL Collaboration],
  Eur.\ Phys.\ J.\  C {\bf 12}, 1 (2000)
  [arXiv:hep-ex/9904015]; \\
%
  G.~Abbiendi {\it et al.}  [OPAL Collaboration],
  Eur.\ Phys.\ J.\  C {\bf 11}, 619 (1999)
  [arXiv:hep-ex/9901037]; \\
%
  R.~Akers {\it et al.}  [OPAL Collaboration],
  Z.\ Phys.\  C {\bf 67}, 555 (1995).
%
\bibitem{hera}
  M.~Turcato  [ZEUS and H1 Collaborations],
  Nucl.\ Phys.\ Proc.\ Suppl.\  {\bf 162}, 283 (2006); \\
%
  A.~Aktas {\it et al.}  [H1 Collaboration],
  arXiv:hep-ex/0703004; \\
%
  S.~Chekanov {\it et al.}  [ZEUS Collaboration],
  Eur.\ Phys.\ J.\  C {\bf 44}, 463 (2005)
  [arXiv:hep-ex/0501070].


\bibitem{Bfac}
Y.~Enari  {\it et al.} (Belle Collaboration),
Phys.\ Lett.\ B {\bf  622}, 218 (2005);\\
  Y.~Miyazaki {\it et al.}  [BELLE Collaboration],
  Phys.\ Lett.\  B {\bf 648}, 341 (2007)
  [arXiv:hep-ex/0703009]; \\
%
B.~Aubert  {\it et al.} (BaBar Collaboration),
Phys.\ Rev.\ Lett.\ {\bf 98}, 061803 (2007); \\
B.~Aubert {\it et al.}, (BaBar Collaboration),
Phys.\ Rev.\ Lett.\  {95} (2005) 041802, 
[arXiv:hep-ex/0502032].



\bibitem{meg}
The MEG Collaboration, http://meg.web.psi.ch/



\bibitem{pdg}
Particle Data Group, Review of Particle Physics, J. Phys. {\bf G}:
Nucl. Part. Phys. 33, 1 (2006).

\bibitem{g_mu}
W.J.~Marciano and A.~Sirlin, Phys. Rev. Lett {\bf 61}, 1815 (1988);\\
T. van Ritbergen and R.G.~Stuart, Phys. Rev. Lett {\bf 82}, 488 (1999).


\bibitem{vud}
J.C.~Hardy and I.S.~Towner, Phys. Rev. Lett {\bf 94}, 092502 (2005) 
[nucl-th/0412050];\\
G.~Savard {\em et al.}, Phys. Rev. Lett {\bf 95}, 102501 (2005).


\end{thebibliography}
\end{document}